\documentclass[letterpaper,twocolumn,10pt]{article}
\usepackage{usenix2019_v3}

\usepackage{tikz}
\usepackage{amsmath}
\PassOptionsToPackage{hyphens}{url}
\usepackage{url}
\usepackage{hyperref}
\usepackage{xcolor}
\usepackage{enumitem}
\usepackage{booktabs}
\usepackage{filecontents}
\usepackage{mdframed}

\usepackage{titlesec}
\titlespacing{\subsection}{0pt}{1.5ex}{.7ex}
\titlespacing{\paragraph}{0pt}{.25em}{1.5ex}
\titlespacing{\subparagraph}{\parindent}{1.9pt}{1.5ex}
\usepackage[normalem]{ulem}

\usepackage{etoolbox}
\AtBeginEnvironment{quote}{\vspace{-1ex}}
\AtEndEnvironment{quote}{\vspace{-1.25ex}}

\newcommand{\new}[1]{{\color{black}{#1}}}
\newcommand{\camera}[1]{{\color{black}{#1}}}

\begin{document}

\date{}

\title{\Large \bf Why Older Adults (Don't) Use Password Managers \thanks{This article will be presented at the the 30th  USENIX Security Symposium (Sec'21), August 11-13, 2021. }}

\author{
{\rm Hirak Ray, Flynn Wolf, Ravi Kuber}\\
University of Maryland, Baltimore County\\
{\tt [hirakr1,flynn.wolf,rkuber]@umbc.edu}
\and
{\rm Adam J. Aviv}\\
The George Washington University\\
{\tt aaviv@gwu.edu}
} 

\maketitle

\begin{abstract}
Password managers (PMs) are considered highly effective tools for increasing security, and  a recent study by Pearman et al. (SOUPS'19) highlighted the motivations and barriers to adopting PMs. We expand these findings by replicating Pearman et al.'s protocol and interview instrument applied to a sample of strictly older adults (>60 years of age), as the prior work focused on a predominantly younger cohort. We conducted $n=26$ semi-structured interviews with PM users, built-in browser/operating system PM users, and non-PM users. The average participant age was 70.4 years. Using the same codebook from Pearman et al., we showcase differences and similarities in PM adoption between the samples, including fears of a single point of failure and the importance of having control over one's private information. Meanwhile, older adults were found to have higher mistrust of cloud storage of passwords and cross-device synchronization. We also highlight PM adoption motivators for older adults, including the power of recommendations from family members and the importance of education and outreach to improve familiarity.
\end{abstract}


\section{Introduction}

Knowledge-based mechanisms have been widely adopted to support security among users.  Strong passwords which are not shared or reused have been recommended since the 1990s~\cite{adamsACM1999}. However, these are known to be challenging to recall, leading to individuals compromising security to achieve memorability.
Multiple studies have attempted to understand the complex factors which play a role in password creation, management and storage~\cite{urUSENIX2012, urUSENIX2015}. Some users' personal password management techniques may involve a trade-off between security and convenience~\cite{stobertSOUPS2014}, whereas others may involve heavy reuse of passwords~\cite{washSOUPS2016}. Understanding password choices and composition also shed light upon the relationship between online privacy behaviors and password strength~\cite{shayCHI2015}.

As a result, security experts often recommend password managers (PMs) as a means of automatic password generation, management and storage. PMs are an effective tool to achieve convenient authentication and improved security when accessing online accounts~\cite{stobertNSPW2014}. PMs come in many forms: standalone PMs, like LastPass or 1Password, that auto-fill, generate, and save passwords; browser-based PMs, like in Chrome or Firefox, that save entered passwords; and operating system PMs, like OSX Keychain, that manage passwords at an OS level across applications, like Wi-Fi passwords.

Given the wide range of choices of PMs, and the widely touted benefits of using these types of technology, Pearman et al. analyzed why users do (and do not) use PMs~\cite{pearmanSOUPS2019}.  The researchers conducted semi-structured interviews with $n=30$ participants split between those who do not use a PM, those who use a built-in browser-/OS-based a PM, and those who use a standalone PM. Participants described their password management techniques, trade-offs between convenience and security in PM adoption, motivations for and barriers against adopting PMs, and uncertainty regarding the source of password-saving prompts on browsers.  

Given that older adults express similar concern towards their digital lives~\cite{eluezeABS2018}, the necessity of convenience~\cite{tsaiEG2015}, and the need for privacy~\cite{quanhaaseJAIST2020} when using technology, PMs may be an effective tool for this user group.  PMs could help address concerns relating to declining levels of physical dexterity and worries relating to cognitive prowess~\cite{caspiAMH2019}.  PM usage could support good security hygiene, while providing a convenient option for interaction---limiting the number of passwords to remember. However, adoption levels of PMs appear to be low overall~\cite{alkaldiUGLASGOW2016}, and studies of PM adoption have seen minimal investigation for older adult groups.

In this paper, we expand upon the research of Pearman et al., which focused on a predominantly younger sample (only one participant was >60 years old). We replicated the study conducted by Pearman et al. with an exclusively older adult sample ($n=26$) to analyze differences in PM usage (and non-usage) with those over 60 years of age. \new{This replication allowed us to directly compare two age groups using the same interview script and codebook.} During the semi-structured interview, participants answered questions relating to authentication habits, password composition strategies, password management techniques, and overall perceptions of and experiences with PMs. We also sought to recruit roughly equal samples of participants who do not use PMs, those who use PMs built into their browsers or operating systems, and those who use separately installed PMs.


Comparing the three PM user groups, we identified a number of differences with the relatively younger sample of Pearman et al. In particular, we highlight a higher number of adoption motivators required by older adults. This includes repeated recommendations from those they trust (close family members), the need to familiarize themselves with the tool's features, and a feeling of urgency to adopt these security tools. However, once motivated to adopt PMs, we see that older adults are tenacious users of PMs and engage in more secure practices (such as using the password generator)  rather than simply convenient ones. At the same time, though, older adults are skeptical of cloud storage of passwords and \camera{do not trust the synchronization process}
 of separately installed PMs. 


There are also noted similarities between the younger group  (from Pearman et al.' study) and older adults (from the study described in this paper). Those who did not use PMs used similar password creation and management strategies based on ad-hoc methods that lead to easily guessable and insecure passwords. This same group, for both younger and older adults, showed little concern of being vulnerable to attacks and did not believe they were engaging in risky behavior. Common barriers towards adoption of PMs were also apparent, such as the fear of a single point of failure, and the importance of having control over one's private information. For the older adults, at least, some of these perspectives may derive from their concern over self-determination in the presence of cognitive decline associated with the aging process.

In summary, we make the following contributions:
\vspace{-.1in}
\begin{itemize}[noitemsep]
\item We present motivations and barriers to adopting PMs among older adults, replicating methods from prior work to allow for direct comparison with younger adults;

\item We analyze the motivations for older adults to adopt and not adopt PMs, such as the cumbersome set-up process, the lack of urgency, and the need for a simpler and convenient method to store passwords;


\item We describe the effect of social influence and self-efficacy for older adults to adopt PMs, such as the impact of family and other advocates;

\item We suggest new techniques for encouraging broader adoption of PMs among older adults.
\end{itemize}


This research suggests that encouraging adoption for older adults, while perhaps more challenging, may lead to strong security outcomes as compared to their younger counterparts.
We suggest focusing on the role of advocacy and education, and particularly, the role of (younger) family members in encouraging older adults to considering PMs. Efforts to improve adoption among younger cohorts may have the secondary effect of also encouraging older adults to maintain good security hygiene. 




\section{Related Work}
\label{sec:related}

\paragraph {Perceptions of security among older adults.}
Older adults are less likely to engage in online activities~\cite{dobranskyPOETICS2016, gellGERONTOLOGIST2015} compared with other age groups.  Challenges can be attributed in part to limited access to Internet services, and inaccessible technologies which have not been designed to account for older adult users' needs.  Older adults are also known to express concerns relating to online security and privacy. For example, Elueze et al.~\cite{eluezeABS2018} found that spam, unauthorized access to personal information, and information misuse were the most pressing issues. Studies have also examined the older adults online activities, highlighting the lack of understanding of online privacy results that can lead to the unintentional distribution of personal information~\cite{nurgalievaPERVASIVE2019} and vulnerability/susceptibility to attacks (e.g. targeted phishing threats)~\cite{linTOCHI2019}. \new{Frik et al.~\cite{frikqualitative} examined the views of older adults relating to information collection, transmission and sharing using traditional ICT and emerging technologies, and found that participants were unwilling to share financial data and medical records.}
    
    
Prior work~\cite{quanhaaseJAIST2020, frikSOUPS2019} also describes older adults' adoption of various privacy protection strategies, such as limiting information shared online, avoiding the use of online services and/or technology, ignoring or deleting online requests, and managing passwords. Evidence of a "privacy divide" was found by Huang et al.~\cite{huangAIST2018}, wherein older adults between the ages of 55-65 were more likely to adopt privacy protection strategies compared to those aged above 65, which may result in older age groups (i.e., those >65) being more vulnerable to online attacks. Das et al.~\cite{dasWAY2019} studied user experiences of two-factor authentication among adults aged 60 or above, and found that older adults did not fully comprehend the benefits. Some participants also reported that they found two-factor authentication to be useless, since they have already adopted a PM and believed that their data could not be breached.

\paragraph {Password managers.}
Password managers (PMs) offer the opportunity for users to centrally store, organize and synchronize passwords across multiple systems and devices. They also come in multiple varieties. Standalone PMs, like LastPass, Dashlane, KeePass, 1Password, StickyPassword, among others, are accessed via the web, browser extensions or apps. Additionally, there are PMs built into many modern web browsers - notably Chrome will save passwords and synchronize them across installation instances. Built-in PMs, like that in Chrome, do not always have password generation features, but this is also becoming more common; Firefox and Safari have such features. Some operating systems also have password management functionality, such as iOS Keystore to save wifi passwords and application passwords, but can also be used to do more general password management. 

Researchers have investigated adoption of PMs~\cite{alsimaniIFIP2010, gastiESRCS2012}, finding low adoption rates~\cite{stobertSOUPS2014, stobertNSPW2014, ionSOUPS2015}, particularly among separately installed PMs. Alkaldi et al.~\cite{alkaldiUGLASGOW2016} suggested that poor advertising and failure to reassure the trustworthiness of PMs are key factors behind low adoption. Fagan et al.~\cite{faganHCI2017} conducted an online survey
and found that PM users valued convenience rather than security 
while non-users found it difficult to trust the security of PMs. 
    
Chiasson et al.~\cite{chiassonUSENIX2006} performed a usability study of two PMs and found that the largest issues is that users have incomplete or incorrect mental models of PMs. Participants were found to experience difficulties "relinquishing control" to the PMs and did not believe that the PMs offered greater levels of security. Control was also found to be an important factor in a study by Karole et al.~\cite{karoleICISC2010} when it comes to password management for non-technical users, who were found to prefer to manage their passwords themselves. \new{Seiler-Hwang et al.~\cite{seilerSIGSAC2019} 
evaluated the usability of mobile PMs, noting three areas for improvement: security, guidance, and integration with external applications.}
    
The auto-fill policies of PMs were examined in a study by Silver et al.~\cite{silverUSENIX2014}.  The researchers identified several risks associated with auto-fill policies which could be mitigated by using PMs to strengthen credential security. Belenko et al.~\cite{belenkoBE2012} analyzed PMs on smartphones, and encouraged encryption of back-up passwords to increase the level of security while using PMs, in the event of attackers accessing the device physically. Using MTurk, Lyastani et al.~\cite{lyastaniUSENIX2018} compared built-in PMs and separately installed PMs. They found that Google Chrome's auto-fill feature encouraged password reuse, whereas LastPass' password generator encouraged stronger passwords that are not reused.

\paragraph {Summary of Pearman et al. methods and findings}
    
Pearman et al. studied password management strategies and motivations for using (and not using) a PM~\cite{pearmanSOUPS2019}. The researchers conducted 30 semi-structured interviews with a roughly even number of stand-alone PM users, OS-/browser-based PM users, and non-PM users. 
Pearman et al. found evidence of a security vs. convenience trade-off, wherein built-in PM users preferred convenience, as opposed to separately installed PM users who preferred security. They also determined a number of factors driving adoption of PMs (such as security, memory issues, and convenience), as well as barriers to adoption for non-PM users (security concerns, trust in the company/tool to not decrypt their passwords). The researchers advocated for tailored advocacy for increased security and PM adoption.

We attempted to replicate Pearman et al.'s method to compare their predominantly younger sample to an exclusively older sample. This included using the same \camera{interview} instrument and codebook to analyze transcripts, allowing for direct comparisons. 
And in many ways, we were able to confirm the findings of Pearman et al. We see key similarities in our sample regarding non-PM users' password creation strategies and the sense that their lack of risky online behavior render them immune from vulnerability. We also see similarities regarding concerns that using PMs may lead to a single point of failure, and the importance of having control over one's own private information. We also find a number of  differences from Pearman et al., specifically their experiences, motivators for and barriers against the adoption of PMs, which are highlighted throughout Section~\ref{sec:results} and summarized in Table~\ref{tab:summary}.



\begin{table}[t]
\scriptsize
    \centering
    \begin{tabular}{r  c  l l}
         \textbf{Part. No.} & \textbf{Age} & \textbf{Gender} & \textbf{PM Usage}  \\
         \toprule
         1 & 71 & Male & Separately Installed PM User \\
         2 & 72 & Female & Non-PM User \\
         3 & 72 & Female &Non-PM User \\
         4 & 73 & Female & Separately Installed PM User \\
         5 & 72 & Female & Non-PM User \\
         6 & 73 & Male & Built-in PM User \\
         7 & 78 & Male & Non-PM User \\
         8 & 71 & Female & Built-in PM User \\
         9 & 61 & Male & Built-in PM User \\
         10 & 66 & Male & Separately Installed PM User\\
         11 & 61 & Female & Built-in PM User \\
         12 & 71 & Male & Built-in PM User \\
         13 & 65 & Male & Non-PM User \\
         14 & 62 & Male & Non-PM User \\
         15 & 68 & Female & Non-PM User \\
         16 & 76 & Female & Built-in PM User \\
         17 & 69 & Male & Built-in PM User \\
         18 & 79 & Male & Built-in PM User \\
         19 & 70 & Female & Non-PM User \\
         20 & 74 & Female & Built-in PM User \\
         21 & 75 & Female & Non-PM User \\
         22 & 75 & Female & Non-PM User \\
         23 & 68 & Male & Separately Installed PM User \\
         24 & 67 & Female & Separately Installed PM User \\
         25 & 75 & Male & Separately Installed PM User \\
         26 & 66 & Female & Separately Installed PM User \\
         \bottomrule
    \end{tabular}
    \caption{Demographic information}
    \label{tab:demo}
    \vspace{-.2in}
\end{table}

\section{Methodology}
We conducted 26 semi-structured interviews with older adults (aged above 60) to understand their password composition strategies, their online habits with authentication, and their opinions and experiences with PMs. We also analyzed responses that spoke to older adult participants' beliefs and concerns regarding securing their online accounts. \camera{For our qualitative findings, we use the terms "a few" as 0\% to 25\%, "some" as 25\% to 45\%, "about half" as 45\% to 55\%, "most" as 55\% to 75\%, "almost all" as 75\% to 99\%, and "all" as 100\% as per Emami-Naeini et al.~\cite{emami2019exploring}. The protocol was approved by our institution's IRB.}

\subsection{Interview Method}
We used the same interview instrument from Pearman et al., and as a semi-structured interview, when answers were unclear, follow-up questions were provided, such as "please explain more" or "could you provide an example?" The general structure of the interview is as follows, and a copy of the survey material can be found in the Appendix.

Participants were presented with a consent form, and were allowed to ask questions to the researchers before the interview began. 
Participants were then asked a series of demographic questions about their age, identified gender, current occupation, level of experience with technology, and a brief description of their biggest online security concern. 

Next, participants were asked a series of questions about their general password usage for online accounts and from which devices they access these accounts, if they manually typed or auto-filled passwords, and if passwords varied across accounts. Participants were also asked what they found easy or difficult about how they managed passwords, and if they ever experienced compromises of their accounts and what they did (or would do) in such situations. 

Participants were then briefed about PMs: "Password managers are tools that can securely handle passwords for you. They can remember your passwords, generate new ones, and even sync them across devices." And following, they were offered descriptions of different forms of PMs, such as standalone PMs, built-in browser PMs, and OS-based managers. They were then asked if any of their personal password management usage fitted these descriptions.

If their usage fitted one of these descriptions, then a series of follow-up questions were asked about why they selected this PM, what they find helpful/unhelpful, if passwords were synchronized, 
management of a master password (if applicable), and if they use password generation tools. Additional questions about security hygiene regarding password changing and data breaches were also included. 

If participants did not use a PM, then they were asked to explain why they do not use one, with additional follow-up questions for more details. They were also asked about awareness of the cost of PMs, or if additional features were available, would this change their mind to adopt a PM. 

Interviews were recorded and then transcribed by a commercial transcription service. Participants were requested not to share any of their passwords or personal information, and if that did happen, these details were removed from the recordings before transcription. Interviews were conducted in person prior to COVID-19 lockdown, and by phone/video conference afterward. Each interview took approximately 30 minutes to conduct for those who did not use a PM and 60 minutes for those who did. Interviews were mostly conducted by the primary researcher, a second researcher did assist with a few.





\subsection{Analysis Methods}
\label{sec:meth:analysis}
Based on participants' responses to general questions regarding their online accounts, passwords, and password management techniques, they were categorized into three groups. This comprised of 10 non-PM users, 9 built-in PM users, and 7 separately installed PM users. These quantities are similar to those of Pearman et al.'s study with a younger population. 

Analysis involved thematic coding of interview transcripts using the codebook provided by Pearman et al.'s codebook~\footnote{Available at \url{https://osf.io/6u7m8/}}. \camera{We chose to use Pearman et al.'s codebook instead of developing our own since their interview protocol was closely replicated and their codebook was checked for reliability (Refer to Section 4 for further details).} We then analyzed the identified themes for the older adult sample to draw conclusions on their behavior by comparing the codes for the three users groups (PM users, OS-/browser-based PM users, and non-PM users). We then compared the identified themes to those presented by Pearman et al. in their description and discussion. We also considered situations where we could not find a reasonable code from Pearman et al.'s codebook. In such situations, we developed a new code and considered this a marker of differences between age groups. We only identified two additional codes related to master password composition and perceived storage of passwords in PMs, which are discussed in Section~\ref{sec:results}


As a reliability check on our coding process, a second, reliability coder used Pearman et al.'s codebook, plus our additional codes, to thematically code 20\% of the transcripts. The reliability coder met with the primary coder to resolve any differences in applying codes. Following, the primary coder updated any codes based on those discussions. The Cohen's $\kappa=0.75$ was achieved for the 20\% sample of coding, suggesting substantial agreement in applying the codebook. 
We did not validate the codebook itself, as this was already done by Pearman et al., who reported a high-agreement inter-rater reliability (Cohen's $\kappa=0.84$). 




\subsection{Recruitment}
\camera{While Pearman et al. recruited through online venues such as Facebook, Craigslist and Reddit, combined with offline strategies such as posting flyers on bulletin boards without adopting snowball sampling methods,} we recruited participants from two state-operated senior centers for in-person interviews and through snowballing following COVID-19 restrictions for phone/Skype interviews. 
%
We found it difficult to recruit standalone PM users who were over the age of 60, which may be anecdotally informative about PM adoption in this group, and so we used purposive sampling to counterbalance. Our final recruitment is similar in size to that of Pearman et al.'s samples. Participant demographics are provided in Table~\ref{tab:demo}.

Participants were eligible for the study if they were over the age of 60 and maintained at least two online accounts. Participants were informed that they were not required to provide any personal information about those accounts, nor their specific passwords. 
In-person participants received \$10 for their participation. It was, unfortunately, not possible to compensate remote participants due to our institutions' policies. Participants were informed of the latter in advance of scheduling their remote interviews.



\section{Limitations}
\label{sec:limit}

When examining issues of security and privacy, some participants may be affected by social desirability bias, wherein they feel obligated to express good security behavior they do not actually practice. 
We used follow-up questions and probes in instances where participants provided brief responses to better understand their true behaviors and practices. 


The outbreak of COVID-19 led to a series of restrictions on recruitment. As a result, we transitioned from conducting interviews from in-person to virtual. We did not find substantive differences between interviews conducted by phone/Skype as compared to in-person.
We were also challenged in recruiting older adult participants that used separately-installed PMs, which may anecdotally suggest issues with adoption, but without a broader survey, we cannot be certain. This led us to conduct purposive sampling via snowballing, which could lead to less representativeness in the sample of separately-installed PM users. The opinions held by this group still offer key points, particularly in comparison to Pearman et al.'s study, as they had a similar distribution of participants.


Finally, \camera{we did not re-interview} younger adults but instead relied on previously published material from Pearman et al. \camera{As described in Section~\ref{sec:meth:analysis}}, we also did not develop our own codebook, instead relying on the previously published one. As the prior work is very recent with a comprehensive, publicly available dataset and codebook, we are confident that our methods match those of Pearman et al. and our results are comparable. Further, the codebook IRR reported by Pearman et al. ($\kappa=0.84$) shows high agreement, as does our own application of the codebook ($\kappa=0.75$).





\section{Findings}
\label{sec:results}

\newcommand{\tabitem}{~~\llap{\textbullet}~~}
\begin{table*}[t]
\scriptsize
    \centering
    \resizebox{1.02\linewidth}{!}{%
    \begin{tabular}{p{5.6cm}  p{5.6cm}  p{5.6cm}}
         \textbf{Similarities} & \textbf{Unique To Older Adults} & \textbf{Unique To Younger Adults (from Pearman et al.)} \\
         \toprule
         \tabitem Ranked financial accounts as more important to protect than other accounts. & \tabitem Non-PM users mentioned that passwords were rarely reused. & \tabitem Non-PM users admitted to heavy reuse of passwords.\\
         \tabitem Non-PM users used specific words in their passwords. & \tabitem Non-PM users were unwilling to pay for PMs. & \tabitem Non-PM users were open to trying built-in PMs.\\
         \tabitem Non-PM users were concerned about a single point of failure. & \tabitem Non-PM users valued having control over who has access to their passwords. & \tabitem Non-PM users valued having control over how their passwords are organized.\\
         \tabitem Built-in PM users were concerned about others having access to their passwords. & \tabitem Non-PM users felt that PMs would not be required since they are unlikely to create more passwords at their age. & \tabitem Non-PM users felt that their accounts were not important enough to require a PM. \\
         \tabitem Built-in PM users and separately-installed PM users liked the auto-fill feature and the convenience of not typing in passwords. & \tabitem Built-in PM users did not express difficulties in password management. & \tabitem Built-in PM users expressed concerns about being unable to update and view all their saved passwords. \\
         \tabitem Separately-installed PM users did not completely trust PMs to always remember their passwords. & \tabitem Built-in PM users were aware of the benefits of separately-installed PMs but felt that the set-up process would be too cumbersome. & \tabitem Built-in PM users were unaware of certain features and benefits in separately-installed PMs.\\
         \tabitem Separately-installed PM users felt that PMs removed the need to memorize passwords. & \tabitem Built-in PM users did not trust separately-installed PMs to be invulnerable. & \tabitem Built-in PM users did not explicitly express any skepticism about the security of separately-installed PMs.\\
         \tabitem Users who adopted separately-installed PMs were motivated by their desire for better security. & \tabitem Master passwords were composed of information that was personal to the user. & \tabitem Master passwords were composed of nonsensical pass-phrases or were randomly generated.\\
          & \tabitem Separately-installed PM users were satisfied with the password generation feature and overall experience of using a PM. & \tabitem Separately-installed PM users found the password generation feature to be inconvenient and would instead engage in risky behavior by re-using older passwords. \\
          & \tabitem Separately-installed PM users expressed distrust towards cloud storage and synchronization of passwords. & \tabitem Separately-installed PM users preferred cloud storage and expressed frustration when the lack of cloud storage hindered their ability to access passwords on other devices. \\
          & \tabitem Separately-installed PM users were recommended to adopt PMs by their family members. & \tabitem Separately-installed PM users were recommended to adopt PMs by staff at their workplace and on online forums. \\
         \bottomrule
    \end{tabular}
    }
    \caption{Summary of similarities between findings from Pearman et al. and our own study.  Findings unique to older adults and those unique to younger groups are also described.}
    \vspace{-.2in}
    \label{tab:summary}
\end{table*}


We follow a similar outline Pearman et al. in presenting our findings. We first focus on the older adult participants and then offer a comparison to prior work.

\subsection{Password Habits}

\paragraph{Prioritizing security of financial accounts above others, and variance in password strategies among groups}
Among older adult participants, separately installed PM users owned more accounts (greater than 150) than non-PM users (less than 50).  Most participants, regardless of experience with PMs, echoed that financial accounts were the most important to protect, as threats to these could cause the most irreparable damage compared to other \camera{sites}. Six participants described a hierarchy of sites where precautions should be taken (e.g., Financial sites > Social Media sites > Casual Media sites (Netflix, news websites).

Most of our participants were aware of the concept of "strong passwords" encompassing an array of letters, numbers and/or special characters.
However, the three groups showed differences in strategies of password composition. Non-PM users tended to use phrases and words of personal significance for their passwords. Built-in PM users  followed a specific pattern, which consisted of a set of characters or numbers which they would move around to generate passwords. Separately-installed PM users generally used a completely random set of numbers and letters, often created by the password generation feature in their password managers. Participant P14, a non-password manager user, said,
\begin{quote}
    \small
    \textit{``It depends on the site. For Amazon, it would be something like `Shop' then a symbol/special character and then a number. I relate it to the site in some way.``}
\end{quote}
On the other hand, password manager user P04 said,
\begin{quote}
    \small
    \textit{``It should be at least 8 characters with a combination of numbers, symbols and letters. And it shouldn't make sense, so just a random combination.``}
\end{quote}

Examples of innovative strategies to generate strong passwords, included romanizing information from words in other language scripts (e.g., Hebrew, Arabic), and interspersing with other characters.  One participant mentioned that this made his password "more random" and more difficult for attackers to guess. Of course, this reflects a perception. Threats to passwords are targeted, rather then based on complex and automated guessing algorithms.  PM users with technical backgrounds preferred using the PM password generation feature, stating that these technologies 
"adjust to the requirements of password generation and its level of sophistication."

\subparagraph{\em Comparison to Pearman et al.} Participants also described a \textit{similar} ranking of account importance in Pearman et al., placing financial accounts higher to protect than others. One of their participants among non-PM users mentioned a similar strategy to our older adult population, of using words related to specific things (such as "kids", "cities", "names"). They also reported owning a \textit{similar} number of online accounts, as well as similar password creation strategies.

\paragraph{Selecting stronger passwords based upon perceived importance of accounts.}
Most non-PM users mentioned that each password was unique, and they were rarely reused. The exception was for accounts termed "casual," where there were instances of repurposing (n=4) or generating a password with roots in an older password (n=2).  However, those participants highlighted not taking risks for those accounts determined "important." where security breaches could prove challenging (e.g., where financial details were present). Some built-in PM users and separately installed PM users shared traits of other users - admitting that passwords for less important accounts shared similarities but more important accounts were unique. This could be attributed to the higher number of accounts that password manager users reported having.

\subparagraph{\em Comparison to Pearman et al.} This was \textit{different} to findings described by Pearman et al., where a number of non-PM users admitted to heavy reuse of passwords, with the exception of one participant who reused substrings for new passwords. A few built-in PM users and separately installed PM users admitted to some reuse, but limited it to accounts which were less important, \textit{similar} to our findings with older adults.

\subsection{Barriers to Adoption Among Non-PM Participants}
\label{sec:barriernon}

\paragraph{Cost-conscious and unwilling to pay for PM software.}
Non-PM users generally responded negatively when questioned about purchasing a PM, expressing that they were cost-conscious, and did not deem a PM to be important enough to pay for. P02 stated,

\begin{quote}
 \small
\textit{``When you're on a fixed income, you're counting your pennies. And you gotta see what's important.``}
\end{quote}

Some non-PM users described multiple bills they were currently paying (utilities, cable, subscriptions) and did not wish to add another item to the list. These participants may be unaware that free PMs exist, which may affect their choice of non-adoption.




\subparagraph{\em Comparison to Pearman et al.} On a similar note to the older adults' negative opinions on the importance of PMs, non-PM users in Pearman's study expressed that they did not deem their accounts valuable enough to require extra security in the form of a PM. Most non-PM users also expressed that they were unwilling to pay for PMs. However, they were more open to trying built-in PMs, or free versions of separately installed PMs. While older adults didn't show any interest in paying for specific features, some participants from Pearman et al. said they would be willing to pay for special features in PMs such as identity theft protection.

\paragraph{Favoring tried and tested methods of password management, along with desire for control.}
Most non-PM users felt that their current method of managing passwords (writing them down) was a safe and easy method, reducing the likelihood of forgetting passwords over time. They expressed the importance of having control over the storage tool used, and showed distrust towards electronic devices which others could control remotely. Participant P05 mentioned,

\begin{quote}
 \small
\textit{``It's simple. I always remember what a book is. And my book is safe. But you can take control of my phone.``}
\end{quote}

\camera{Some} non-PM users expressed mild annoyance about the low portability of writing passwords down but didn't consider this as a priority for password management.

\subparagraph{\em Comparison to Pearman et al.} Among non-PM users in Pearman's study, some participants kept their passwords in a list stored on their mobile device, since they valued portability, and were unaware of other methods of accessing passwords on-the-go. The importance of control was also seen among Pearman et al.'s younger adults, but from a different perspective. They valued the control over how their passwords were organized, and being able to categorize them in specific ways in a notebook.

\paragraph{Incomplete and erroneous mental models of password storage within PMs.}

Mental models were found to vary considerably between different types of \camera{users}.  
When asked about storage of data, most non-PM users were unable to venture a guess as to how passwords are stored in PMs.
One non-PM user 
described, inaccurately,  that passwords were possibly stored as shortcuts, such as keyboard shortcuts.

Safety concerns led most non-PM users to be adamant about keeping their private information inside their homes, local computer storage, or physically among their belongings. As expressed by P22,

\begin{quote}
 \small
\textit{``Local storage is more secure because it adds a dimension of physical security that you can control. It is only in one place.``}
\end{quote}

\subparagraph{\em Comparison to Pearman et al.} A few non-PM users in the study by Pearman et al. expressed concern about using built-in PMs. One non-PM user felt that passwords stored in a built-in PM might be lost due to a memory wipe of the device in her workplace. Both age groups were concerned about the security of passwords in locations out of their control. However, Pearman et al.'s participants were also unsure whether their passwords would be stored correctly.


\paragraph{Worries about dependence on technology to manage passwords}
\camera{Most} non-PM users voiced concern regarding a single point of failure (putting all their passwords in a single location)
, and felt that computers can crash at any moment impacting access to their accounts, or passwords may get accidentally deleted.  Becoming dependent on technology could be problematic. P02 mentioned,

\begin{quote}
 \small
\textit{``What if your computer or something is down and all your passwords are stored and you can't retrieve 'em? ``}
\end{quote}

Other older adults felt that using PMs would be acknowledging their diminishing ability to remember information.
P19 mentioned,
\begin{quote}
 \small
\textit{``It's like, technology is great. But what technology does is [it affects] things that we store in our memory when we put it in technology. Case in point, how many of us remember phone numbers anymore? If we keep relying on technology, we lose the ability to think for ourselves.``}
\end{quote}

\subparagraph{\em Comparison to Pearman et al.} 
Both samples expressed concern about keeping all their passwords in one location and the risks of a single point of failure. Younger adults in Pearman et al.'s sample were also worried about unsolicited individuals getting access to all their passwords in one attempt, while older adults in our study were afraid of putting their trust in the  technology which may fail, rather than their account being "hacked." While some younger adults felt that they were giving up a feeling of control by using a PM, this loss of control appeared more acute for older adults as to them it may suggest a decline in their cognitive abilities, or the loss of a way of doing things that they are sentimental about.

\paragraph{Perceived benefits of convenience and portability do not \camera{outweigh} security concerns.}
The overall perceptions of PMs among non-PM users were negative. While some acknowledged a few benefits of PMs (such as the convenience in portability and some additional features like password generation), they still showed resistance and were adamant against adopting PMs. P13 mentioned,

\begin{quote}
 \small
\textit{``Sometimes I might be annoyed because I would be away from home and I wanted to log in to something and I didn't have the password with me. But I wouldn't use one [password manager].``}
\end{quote}

Their concerns with the security (as well as the price of third-party applications, see above) seemed to outweigh PMs perceived benefits. They also appeared to be fairly satisfied with their current methods of managing passwords, and believed having control over their passwords themselves was important. P15 said,

\begin{quote}
 \small
\textit{``I don't think it [password manager] is as secure as keeping it in my good old address book.``}
\end{quote}

Age also proved to be a factor, as some non-PM users expressed their disinterest in PMs since they would be unlikely to create more passwords. P03 expressed,

\begin{quote}
 \small
\textit{``I'm not going to be making any more passwords now at my age. At this time in my life, I'm not going to have any more accounts than I already have. So I wouldn't really need it.``}
\end{quote}

\subparagraph{\em Comparison to Pearman et al.}  
Some participants from both age groups were reluctant to use a PM. While older adults expressed that they would be unlikely to create more passwords and would thus not require a PM, younger adults in Pearman et al.'s study felt that their accounts were not important enough to require extra security offered by PMs.

\subsection{Experiences using Built-in PMs}




\paragraph{Satisfied with auto-fill functionality when accessing accounts.} 
\camera{Almost all} built-in PM users were very pleased with the auto-fill feature in their browsers and operating systems. While some admitted that a few risks were involved, and showed small concerns regarding the security of passwords stored in PMs, the convenience of this method seemed to outweigh their concerns. A few participants expressed that PMs built into browsers are simple to use, and removes the need to remember passwords themselves.

\subparagraph{\em Comparison to Pearman et al.} This was similar to the built-in PM users' opinions in Pearman et al.'s study, who emphasized liking the auto-fill feature, and enjoyed the convenience of not having to type their passwords.

\paragraph{Limited levels of concern voiced regarding password management by non-PM users.}
Most non-PM users did not explicitly describe difficulties or negative aspects of their password management methods that were covered in the questions asked. However, there may be certain habits that they didn't adopt, such as strong encryption of their passwords. One participant (P9) expressed concern about a third-party gaining access to a browser's passwords.

\begin{quote}
 \small
\textit{``Sometimes I wonder whether my passwords are safe there[in browsers]. What if someone just gets in? ``}
\end{quote}

\subparagraph{\em Comparison to Pearman et al.} While some built-in PM users in Pearman et al.'s study also expressed concerns regarding others having access to their passwords, their worries were directed to other individuals using their devices physically. These younger adults also mentioned concerns about the built-in PMs being unable to update their saved passwords, and being unable to view all their saved passwords.

\subsection{Barriers to Adoption of Separately Installed PMs by built-in PM Participants}

\paragraph{Lack of strong enough incentives to change current habits among built-in PM users.} \camera{Most} built-in PM users felt that their passwords may be accessible to others
to an extent using their current password management techniques when compared to separately installed PMs, but also admitted to not having strong enough incentives to adopt a PM such that they would be willing to change their regular routine.
\begin{quote}
 \small
\textit{``My son has been trying to get me to [adopt a separately installed password manager] for a long time. I have not done it yet. It's a good idea, but I'm not used to doing it. I haven't done it yet because of inertia and laziness. I'm so used to getting up in the morning and automatically logging into Google.``}
\end{quote}
Participant P08 felt a lack of urgency and therefore felt it unnecessary to adopt a password manager,

\begin{quote}
 \small
\textit{``I would have to be convinced that it is really beneficial to me in some way. Nothing has happened to me to motivate me to use a password manager.``}
\end{quote}

Built-in PM users also felt that it “wasn’t worth the hassle” and that it would be too difficult to set-up.  They were highly confident that separately installed PMs are the safest method of storing passwords, but conditioned its convenience on the difficulty level of setting it up.

\subparagraph{\em Comparison to Pearman et al.}
Upon comparing the findings, built-in PM users among both age groups stressed the important of the convenience for storing passwords and were aware that their password habits were risky and better methods exist. Both did not take the necessary steps to improve their situation, but the reasons differed. 

While younger adults in Pearman et al.'s study were \textit{unaware} of certain features in separately installed password managers that may be advantageous (e.g. the password generation feature) 
and thus could not properly reflect on the potential benefits, the older adults in our sample expressed \textit{awareness} regarding the benefits of separately installed password managers but were convinced that the installation and set-up process would take too much effort, even without attempting to install them in the first place. 

\paragraph{Trust in PM technologies influences adoption among older adult PM users.}

Levels of trust using PMs varied across user groups. For example, built-in PM users were generally more skeptical compared to other user groups regarding the security of PMs,
and felt that any system could be hacked into. 

\begin{quote}
 \small
\textit{``They can hack into even the government so what is to say they can't hack into a password manager? ``}
\end{quote}


They agreed that portability of separately installed PMs is a worthwhile feature but mentioned that it may take them a while to develop trust in adoption. The perceived set-up process of separately installed PMs also served as a hindrance for built-in PM users to adopt a separately installed PM; however, none of the built-in PM users actually attempted the set-up process.  They also did not trust separately installed PMs to store passwords in the cloud and expressed a desire to have more control over who has access to their information. These same participants seemed unaware that built-in PMs may also store passwords in the cloud under certain configurations.

\subparagraph{\em Comparison to Pearman et al.}
Some built-in PM users in Pearman et al.'s study did not explicitly mention distrust towards separately installed PMs, which differs from our older adult sample. However, they showed signs of distrust towards built-in PMs, and confusion regarding the storage of their passwords in browsers. This confusion led them towards losing their trust in the reliability of these built-in PMs, and resorting to other insecure methods.



\subsection{Experiences using Separately Installed PMs}

Among PM users, 1Password was the most popular application (n=4), followed by LastPass (n=2), and DashLane (n=1). Three participants also kept back-ups of passwords in address books. Most PM users highlighted satisfaction with their PMs. 

\paragraph{Master passwords were composed of personal information.} 
Most PM users had developed master passwords for their PMs, which were retained in memory without any further digital or physical record of them. To ease the process of recalling information, five participants  revealed that their master password was memorable, composed of information that was personal to them or someone close to them (e.g., former phone number, mother’s middle name, spouse’s maiden name), which was combined with other numbers and special characters which also carried meaning to them (e.g., birth dates etc.).  A sequence of random characters would be more complex to remember, and could result in time being spent to recall or reset should it be forgotten. 
As described by P25,

\begin{quote}
 \small
\textit{``It's something that I would remember. I memorized it. No memory aid. No physical copy of it. It is just my wife's name with numbers and special characters. It is not gibberish 
More of a phrase really.``}
\end{quote}

\subparagraph{\em Comparison to Pearman et al.} Prior work showed vast differences in master password composition and management. Some participants mentioned using pass-phrases like movie quotes or sentences that didn't make any sense, whereas others used passwords which were randomly generated. Since some of these passwords were difficult to memorize, they would often resort to keeping written copies, or saving them as email drafts. Older adults in our study, in contrast, explicitly selected master passwords they can manage without aides; not to suggest these are more/less secure than the master passwords of the younger sample.

\paragraph{Extensive use of automatic password generation functionality among certain separately installed PM users.}
Four separately installed PM participants described using the password generation function extensively.  Three did not (of which one was not aware it existed). Of the four who use it, three had no complaints and were highly satisfied, as it helped mitigate the need to think about how to compose a strong password. However, one participant felt that the functionality allowed for setting lengthy passwords, describing that it  "went too overboard."  That participant adjusted the slider to choose the number of characters. Despite this, she also mentioned that an average of 75-90\% of all her passwords were generated using this feature. None of the built-in PM users reported using a password generator, nor any of the non-PM users. 

\subparagraph{\em Comparison to Pearman et al.} While all separately installed PM users in Pearman et al.'s study reported using password generators, \camera{some} were not satisfied with the experience and found it inconvenient, and would sometimes engage in risky behavior by reusing older passwords instead. 
This could suggest that older adults view the password generation feature as a more convenient and positive feature.

\paragraph{Appreciative of specific PM features.} PM functionality favored included the historical record kept of the password and the Face ID feature (n=5), the auto-fill capability (n=2), and being able to change length and type of characters of their password with simplicity and ease (n=2).  In terms of downsides, one participant mentioned that they would prefer the customization settings to be more visible, reducing the time to get going with the software (e.g., to adjust the frequency of logins using the master password). Two participants were not entirely comfortable with the stability of the system. As stated by P25 (although, we are unsure exactly how this occurred),


\begin{quote}
    \small
    \textit{``One time it just stopped working. It was gone. I lost the whole account.``}
\end{quote}

A couple of PM users explicitly mentioned the simplicity of the set-up, learnability and the ease of use. Most PM users described their positive experience using the password generation feature. A few participants liked the organization and consolidation of data entry and the ease of data migration between devices. Some participants had no complaints regarding any features of the password manager. One participant (P01) did not like the updated appearance of the user interface, and mentioned that the previous interface was easier to navigate. Another participant (P10) felt interrupted by the password manager whenever they visit a new website.

\begin{quote}
 \small
\textit{``They always interrupt me when I’m on a new site. They always offer to generate a new password for me. I don’t want to.``}
\end{quote}
Despite interruption, P10 still used password generation.

Most separately installed PM users appeared vaguely aware but did not use the PM's dashboard function which evaluates the strength of current passwords.
P24 expressed indifference, 

\begin{quote}
 \small
\textit{``It always tells me the password I have chosen is not that strong. But I don’t care. I don’t pay attention. According to them, most of my passwords are medium strength, whatever that is.``}
\end{quote}
P23 felt that his encrypred passwords were inaccessible to the dashboard and thus did not pay attention to it,

\begin{quote}
 \small
\textit{``I don’t think the dashboard even knows what my passwords are, because they are encrypted. The company 1Password cannot decrypt them.``}
\end{quote}
This also demonstrates some confusion regarding how data is handled by separately installed PMs, where only the PM clients (with access to the master password) can decrypt the passwords, and thus the dashboard information is actually generated locally. 

\subparagraph{\em Comparison to Pearman et al.} Similar to our study, some participants from Pearman et al.'s study were very pleased with the auto-fill feature of PMs. However, some of Pearman's participants were also displeased about the lack of certain features, such as being unable to enter their long passwords into non-compatible devices and a few websites, which was not mentioned by participants in our study. This could be attributed to older adults using a smaller range of devices and websites than younger adults.

\paragraph{Factors motivating participants to pay for PMs vary depending on extended functionality available, and recommendations from experts.}
Five participants used the paid version of their password managers and two participants used the free version.
One participant who used the paid version believed that the people who designed it should be rewarded, as it offered enough features to feel secure. Another participant who used the paid version said that they wanted the data migration feature which was in the paid version.
Those who used the free version said they would only be willing to use the paid version if recommended by somebody who is an expert, and would need to see a demonstration.
Participants paid \$25-\$30  a year to purchase or maintain subscriptions.

\subparagraph{\em Comparison to Pearman et al.}
Separately installed PM users in Pearman et al.'s study, who used a free version of the PM, mentioned that they would only pay a fee if the tool was very secure and user-friendly. Those who were willing to pay for a PM expressed that they would pay less than \$5  a month.

\paragraph{Experience of setting-up PMs varies, which may limit usage.}

Some participants mentioned that PMs were very easy to set-up and there were clear instructions available. The experience was described as friendly and simple to follow. P25 expressed his contentment with the ease of installation and user-friendly guidelines for adopting PMs,

\begin{quote}
 \small
\textit{``It's really simple, easy and friendly. The website had a list of instructions on what to do. The instructions were very clear, and if I followed the steps as it was written, it was no problem. It just works. I could see problems faced by those who don't understand technology.``}
\end{quote}

On the other hand, some participants mentioned feeling overwhelmed by the number of options and settings while installing their PMs. Participant P01 expressed,

\begin{quote}
 \small
\textit{``The password manager would drive you crazy with the number of features it showed. I had to pick out the ones that I need.``}
\end{quote}


\subparagraph{\em Comparison to Pearman et al.} We see similar opinions in Pearman et al.'s study. While most younger adult participants did not face many issues, a few separately installed PM users found the experience cumbersome and unsatisfactory, sometimes resorting to reusing older passwords instead. Some positive sentiments about the set-up process may also be affected by the fact that most older adults had simpler set-ups, exclusively used the PM on one platform, either mobile or desktop, and did not trust synchronization features (see below). 

\paragraph{Some distrust towards security of storing data in the cloud.}
Aside from two separately installed PM users who use the PM on a desktop device, the remaining five participants described using PMs exclusively on their mobile phones. Participant P25 mentioned the reason for this being that he does not trust cloud storage to securely store his passwords,
\begin{quote}
 \small
\textit{``Because I don’t want to go into the cloud. If I use it on the desktop, I need to consent to cloud storage.``}
\end{quote}
This reflects a recurring misconception about password storage in PMs which will store passwords encrypted in the cloud, even when only installed on a single device. 

Another participant (P23) justified not using the PM on his laptop, claiming it to be unnecessary because his laptop is heavily encrypted.
\begin{quote}
 \small
\textit{``I don't use a password manager on my laptop because it is always with me. Everything on my laptop is encrypted.``}
\end{quote}
This demonstrates a second common confusion;
threats to passwords often occur externally, due to data breaches, not from insecurity of personal devices. 


\subparagraph{\em Comparison to Pearman et al.}
\camera{Some} separately installed PM users from Pearman et al.'s study specifically enjoyed using a desktop client of their PMs and expressed frustration when a lack of cloud storage hindered their ability to access passwords from their mobile devices, which differed from older adults' distrust of cloud storage. This discrepancy with the older adult sample may derive from misconceptions about the way in which passwords are stored and synchronized, as well as misunderstandings about local security risks (such as access to a laptop) and remote access security risks (such as access to an online account).

\paragraph{Skepticism regarding synchronizing passwords.}
While participants with separately installed PMs did not use these technologies across devices (see above).  They also shared dislike and distrust with synchronizing passwords, describing synchronization as being an insecure process which increases their exposure to breaches. P04 expressed her concerns along these lines.
\begin{quote}
 \small
\textit{``If they claim to be syncing passwords across all my devices, that means they are storing them somewhere outside my apartment.``}
\end{quote}
Again, this also expresses a misconception regarding how passwords are managed by PMs, which have encrypted cloud backups to provide synchronization. 

A similar sentiment was expressed by P25, who was afraid of others gaining access to his mobile phone, which he believed would provide access to the password storage. 
\begin{quote}
 \small
\textit{``I don't want that. It is too easy to leave your phone on somewhere. Someone scrolls through it and forwards it to their email. For security reasons, I wouldn't want that.``}
\end{quote}
However, the PM on the phone would still need the master password (or a biometric) to access the passwords.

One participant (P01) did acknowledge the convenience of syncing, and felt that there are no consequences if the passwords are strongly encrypted in the first place. He felt that synchronization of passwords and sharing of accounts is perhaps manageable as long as control and ownership of the account remains intact.

\begin{quote}
 \small
\textit{``I am hypocritically comfortable with sharing Netflix. But the way it is shared I think the account owner can maintain control of the password. It would be up to the account owner to log in. The password should be protected by the owner. Control is important.``}
\end{quote}

Another participant (P24) also valued the syncing feature. She felt that it is convenient and that there are no consequences as long as it is strongly encrypted. She also felt that a backup would be useful if her phone were damaged or lost.

\begin{quote}
 \small
\textit{``I access the same thing across different devices. There aren't any consequences if the part involved in the sync is strongly encrypted. If I lose my passwords, I would be sunk. I have too many accounts. I would hate to have to [go] back and say I Forgot My Password for every account. For me, it's peace of mind.``}
\end{quote}

\subparagraph{\em Comparison to Pearman et al.}
The overall dislike of synchronization of passwords differed from that of younger adults in Pearman et al.'s study, where multiple participants expressed synchronization as an important feature. Without synchronization, some participants in Pearman et al. even resorted to emailing their passwords to themselves. In contrast, our older adult participants view sync-ing as a potential threat to the security of PMs, particularly because they inaccurately believe that the passwords would only be stored locally if they choose not to sync. They also do not see this as an impediment to PM convenience, perhaps because they have fewer devices and access fewer accounts.

\paragraph{Overall satisfaction and confidence in using PMs among participants who utilized them.}
Aside from one older adult password manager user in our study, the remaining participants described being satisfied with their experience of using password managers. They described multiple features that they felt confident using, such as the password generator, and the auto-fill feature.

\subparagraph{\em Comparison to Pearman et al.}
Younger adults from Pearman et al.'s study appeared to offer more complaints about PMs. This could be because the older adults use PMs less often in fewer contexts (e.g., mobile only) and utilize fewer features (e.g., no synchronization).
It may also reflect that the older adult sample, having gone through greater effort to adopt PM's than younger adults, are attempting to justify that experience with stronger feelings of satisfaction. 


\paragraph{Concerns regarding PMs' abilities to maintain passwords data over time.}
Most of the separately installed PM users did not fully trust PMs to remember their passwords despite high satisfaction with PMs. Some of these participants were simply skeptical of having complete faith in PMs, and one participant's distrust derived from a bad experience of their PM disappearing from their device.
P10 highlights errors which may occur during data migration, and the importance of maintaining a back-up of passwords, since they did not complete trust in the PM's ability to store passwords.

\begin{quote}
 \small
\textit{``There can always be an error in the program or when you transfer to a new phone. It’s only human. Some human set it up and anyone can make a mistake. That's why I keep a back-up. No matter what DashLane does, I still have my contact list. I guess the fact that I have a back-up says I don’t trust it.``}
\end{quote}
One participant completely trusted his PM.  In follow up questions, he revealed that this trust stems from a sense of control over which passwords are stored and which are not, and that they do not change without direct action.  Participant P04 said,
\begin{quote}
 \small
\textit{``I’m expecting that the password manager will only change [the password] if I change the password. I am in control when it is changing.``}
\end{quote}

\camera{Some} separately installed PM users felt that errors occurred (a form of mistrust) and kept a back-up in case of emergencies, such as system crashes or problems with their devices.
However, when it came to security, most of the separately installed PM users trusted their PMs to keep their information safe from external attacks and valued the additional security.

\begin{quote}
 \small
\textit{``I think PMs protect my passwords from external threats. That's part of why it's there. I believe they [passwords] are safe and secure through being encrypted inside of the program.``}
\end{quote}

\subparagraph{\em Comparison to Pearman et al.}
Similar trust issues with the PMs' ability to remember passwords were seen in Pearman et al.'s study, where participants reported instances where the PM would incorrectly save usernames and passwords. Instances were also described where a participant did not trust their PM to submit completed credentials while logging into online accounts, as errors by the PM were made in the past.

\subsection{Adoption Motivators to use Separately Installed PMs}

\paragraph{Needing PMs due to the volume of passwords which need to be remembered.}
Most of the separately installed PM users felt that they had too many passwords and wanted an easy way to store them. Some also desired greater levels of security. One participant mentioned that the user reviews of the PM on the App Store were very high, which encouraged him to install it on his mobile device.

\subparagraph{\em Comparison to Pearman et al.}
Similarities were seen with Pearman et al.'s study, where most separately installed PM users described using PMs as a better way to store passwords, removing the burden of memorizing multiple passwords, and having to manually enter passwords when accessing systems. 

\paragraph{Recommended to use PMs by family members.}
%
Most separately installed PM users mentioned that a family member recommended they adopt a PM. One participant (P24) was also given guidance on how to create better passwords. 

\begin{quote}
 \small
\textit{``My son recommended it to me...I had a brief discussion with him on how to create my passwords and he helped me.``}
\end{quote}

Some participants were advised on password generation guidelines by experts in the field whom they trusted. As expressed by participant P24,

\begin{quote}
 \small
\textit{``It would have to be somebody who had, you know, in-depth knowledge, not just a lay person
knowledge, but in-depth knowledge of Internet security issues.``}
\end{quote}

\subparagraph{\em Comparison to Pearman et al.}
Some separately installed PM users in Pearman et al.'s study received recommendations from different sources.
This included being recommended by staff at their workplaces, and on online forums, like Reddit.

\paragraph{Security benefits of PMs outweighing benefits of alternative solutions among separately installed PM users.}
Five of the seven separately installed PM users believed their passwords to be secure in their PM due to them being built by “skilled programmers” and being heavily encrypted. Participant P10 mentioned,

\begin{quote}
 \small
\textit{``Since I am using it only on the PC, all the stuff is stored in the PC and nowhere else, and maybe they are encrypted. Maybe I’m wrong.``}
\end{quote}
On the other hand, two participants were skeptical and did not believe PMs to be absolutely secure. Participant P01 said,

\begin{quote}
 \small
\textit{``There is always a chance that something might go wrong. You learn not to feel too good about these things. Based on what I read, what experts say about it, I feel comfortable in that. But then again, I don't know how secure it is. I cannot say with 100\% certainty that it is.``}
\end{quote}

Participant P23 believed that the security of his PM does not matter, since attackers would need to access to his mobile phone first. He believed that attackers would need to physically obtain his mobile phone to get access to his passwords.

\begin{quote}
 \small
\textit{``No, because it is only on my device. It is unlikely someone can get my phone and then [get] into my password manager.``}
\end{quote}
This is likely a misconception. Passwords are backed-up (and encrypted) on cloud services based on the master password. 

\subparagraph{\em Comparison to Pearman et al.}
Similarities were seen with Pearman et al.'s study, where \camera{most} participants expressed a desire for increased security to be an adoption motivator for PMs and were satisfied with the encryption used by PMs. One separately installed PM user believed he was 
able to store passwords that were not vulnerable to "dictionary attacks." However, some participants resorted to insecure practices when their PMs did not function as intended, which involved reusing older passwords instead of randomly generating new ones. This differed from older adults' positive experiences with password generation using PMs.

\section{Discussion}
\label{sec:discussion}


When comparing our results to Pearman et al. in Section~\ref{sec:results}, there are  commonalities: valuing secure access to financial accounts above other types of online \camera{accounts}, concerns such as a fear of a single point of failure (e.g., losing access to all passwords stored in one place), and the importance of having control over one’s private information.  In terms of differences, master password management strategies were found to vary.  Older adults were found to have a higher mistrust of cloud storage of passwords and cross-device synchronization.  

\camera{We also observed various alternatives towards password management by non-PM users. Multiple participants preferred writing their passwords in an address book, allowing them to maintain an organized and portable record of all their passwords. This certainly encourages using and eases the use of different passwords for different sites, and these older adult users may not need to adopt a PM for their security needs. We believe that PMs offer a benefit to many older users, but we do not argue its a panacea or the right solution for everyone. It likely offers significant improvements over current methods, especially those without meaningful password manage strategies, but as we identify here, the burden to adoption may be too high for many older adult users.}

In this section, we broadly 
discuss the barriers of adoptions we observed, as well as how these results could be applied to encourage wider usage of PMs \camera{for older adult users who would find benefit from adopting a PM}. Finally, we discuss lessons learned from the user experience of PMs and how to encourage more effective use of PMs.

\subsection{Barriers to PM adoption in Older Adults}

\paragraph{Time management and disruptions.}
Older adults may be more sensitive to their management of time online, for example, stating things like Facebook is a "time-waster," which may decrease motivations to invest time in adopting a PM.
A large number of built-in PM users expressed that setting up and installing a stand-alone PM would require a large amount of effort and would be “too much of a hassle," all the while aware of potential benefits with regards to security.
Some older adults simply expressed hesitation in disrupting their current privacy behaviors, which were deemed sufficient.

\paragraph{Effect of technology on memory and control.}
Users' mental models of PMs are known to be incomplete or inaccurate~\cite{chiassonUSENIX2006}. 
The importance of control was mentioned by older adults and their need to feel less dependent on technology. Their cognitive process related to memory was found to be highly important and connoted negatively on their physical decline. Non-PM users narrated anecdotes of being able to remember phone numbers of close friends, and relying on technology for these purposes would inadvertently affect their ability to do the same. Concerns were also raised on an overarching issue with younger generations' addiction and dependency on mobile devices, and their impact on cognitive abilities. Relying on technology to remember information may also be a signifier of increased age, and a subconscious fear of losing the ability to recall information.


\paragraph{Lack of self-efficacy.}
Some older adults' inexperience and low confidence with technology led them to quickly blame themselves when challenges emerged. This may be a consequence of perceptions of a digital divide among older adults.
Small errors quickly compound, leading to self-blame and eventually abandoning adoption, or never attempting in the first place. Some PM users were quick to believe that issues are brought about by their own actions and concerns, rather than the usability or the software itself. 



\paragraph{Lack of trust.}
\new{Alkaldi et al.~\cite{alkaldiUGLASGOW2016} found that a significant barrier towards adoption of PMs was the lack of reassurance to potential users about their trustworthiness. We see similar signs of mistrust among older adults, especially when it comes to storage of their passwords. Older adults showed concern about cloud storage and who has access to passwords stored in the cloud. A higher level of transparency showing users how secure their passwords are (when stored in the PM) could help towards alleviating their concerns and increasing levels of trust, which is also shown in Alkaldi et al.'s study.}

\subsection{Encouraging PM Adoption in Older Adults}

\paragraph{Advocacy from family members.}
Advocates have played an important role in supporting security among a wide range of users ~\cite{wolf2019pretty, haney2018s}.  Advocacy organizations such as AARP have begun informing older adults on interventions to use to support security, through targeted web sites and podcasts \cite{AARP2020}. Encouraging advocates to highlight the effectiveness of PMs, and to offer on-going support during the process of set-up and using these technologies, can offer promise to older adults.

We find that advocacy is also important, particularly from family and close friends. Many participants described their adoption of PM was driven by suggestions and advice from loved ones. For example, participant P24 stated that she started "using LastPass because her son showed it to her." He had used the technology successfully, as he was an advocate.  Multiple built-in PM users said that close friends and family had brought up PMs in conversation, which allowed them to consider trying these out. This implies that encouragement and advice from known members (e.g. family, close friends) are valued by older adults, and conversely, family members seem to be concerned about the security of older adults’ private information. As such, encouraging adoption of PMs of younger adults may in turn increase older adult adoption as these users become advocates to their close friends and, particularly, their older family members. 


\paragraph{The role of education and outreach.}
Some non-PM users were aware of their risky behavior and insecure methods, but showed little motivation or any sense of urgency in adopting better password management techniques. We have seen evidence of urgency being a wake-up call for older adults, wherein social isolation (brought about by COVID-19) has encouraged older adults to move towards adopting online technology~\cite{poon2020citylab}. 
\new{The lack of a sense of urgency shown by older adults was also seen among participants in a study by Aurigemma et al.~\cite{aurigemmaIBISPR2017} with undergraduate students, which may suggest an overlap between students and older adults' lack of a sense of urgency.}

Education and outreach can help older adults better understand the urgency of secure practices. Classes at senior centers and libraries, which have begun to adopt security initiatives to support older adults, could be a vital outlet to disseminate this information. Classes could incorporate the practical application of online security tools (such as PMs) in these classes, while taking into account the mental models of older adults. Older adults could test out PM technologies, which may better address worries regarding the learning curve faced. 
\new{Erroneous and incomplete mental models of how PMs work (For example: encryption, cloud storage, etc.) surfaced multiple times during our interviews. The role of education could help towards correcting these mental models which may impact their decision to adopt PMs in the future.}

\subsection{Design Implications for PM Adoption}

As described in Section~\ref{sec:barriernon}, a few non-PM users showed concern (or confusion) about storing their passwords in password managers and not knowing how or where their passwords are stored (i.e., on the device, in the cloud etc.). To address this, targeted advertising providing clear and visible context for the storage of passwords could potentially help alleviate some of these fears. Modifying PM interfaces to display content which reassures users of the security of their passwords could also prove to be beneficial and invoke greater levels of trust. \camera{ Pearman et al. also suggest emphasizing the sorting and retrieval capabilities of separately installed PMs for those users who value organization of their passwords.} \new{Most built-in PM users liked using the auto-fill feature of browser-based PMs. This feature could be made more transparent in PM interfaces to encourage this group of users.} A few non-PM users also mentioned being required to add extra characters to meet the requirements of some websites, suggesting that their passwords may be too weak to be used. They would often resort to adding characters as instructed by the websites. This could be a good opportunity to provide nudges on these websites, to use PMs for better password generation. This suggestion was also offered by Pearman et al. for a younger demographic, which implies that both age groups could benefit from this. \new{Lyastani et al.~\cite{lyastaniUSENIX2018} also suggested that the existence of password generators is beneficial. Our findings confirm this as well; older adults who used the password generation feature in PMs showed appreciation for its benefits towards creating stronger and lengthier passwords.}

Some built-in PM users felt that the set-up process for PMs was too cumbersome
, even though they were aware of the potential benefits of security associated with PMs. PMs could potentially be designed to allow for a more streamlined set-up process or better demonstrating how this process works as part of promotions. Since built-in PM users mentioned that they heard about PMs on websites, these design implications could also be expressed via third-party websites as methods to improve security hygiene. \camera{While separately installed PM users in our study seemed comfortable with (and aware of) their PM's features, Pearman et al. found that separately installed PM users were unaware of features involving automatic replacement of weak passwords, and they suggest implementing a feature to provide assistance to users in improving existing passwords at the time of PM adoption.}

\section{Conclusion}

In this paper, we describe a study examining the reasons why older adults use (and do not use) password managers (PMs), replicating a protocol used by Pearman et al.~\cite{pearmanSOUPS2019}. We found that opinions regarding online security, PMs, trust and password management creation strategies differed among non-PM users, built-in PM users, and separately installed PM users in the older adult sample. 

Using the same codebook as Pearman et al., we directly compared our older adult sample to Pearman et al.'s predominately younger sample. Older adults express more favorable experiences using PMs. Conversely, they also have a higher mistrust of cloud storage of passwords and cross-device synchronization.  They shared common concerns when it came to the risks of a single point of failure.  

We discuss possible adoption motivators for PMs. Older adults who adopted PM were repeatedly recommended to do so by close family members. These advocates are crucial in encouraging broader adoption, and so actions to improve adoption among younger adults will percolate to the older population as well. Additionally, we identify the role that education and outreach can play to help provide familiarity to PMs, as well as providing more sense of urgency to utilize them by better describing the risks associated with poor password management practices. We also offer design implications for PM adoption targeted towards older adults.

\subsection*{Acknowledgements}
This material is based upon work supported by the National Science Foundation under Grant No. 1845300.  
The authors thank Heera Lee (UMBC) for her help with the research, as well as Pardis Emami-Naeini for shepherding the paper and the feedback from the anonymous reviewers. 

\bibliographystyle{plain}
\bibliography{ref}

\begin{thebibliography}{10}

\bibitem{AARP2020}
{AARP - Privacy and Security}, 2020.
\newblock \url{https://www.aarp.org/technology/privacy-security/}.

\bibitem{adamsACM1999}
Anne Adams and Martina~Angela Sasse.
\newblock Users are not the enemy.
\newblock {\em Communications of the ACM}, 42(12):40--46, 1999.

\bibitem{alsimaniIFIP2010}
Haitham~S Al-Sinani and Chris~J Mitchell.
\newblock Using cardspace as a password manager.
\newblock In {\em IFIP Working Conference on Policies and Research in Identity
  Management}, pages 18--30. Springer, 2010.

\bibitem{alkaldiUGLASGOW2016}
Nora Alkaldi and Karen Renaud.
\newblock Why do people adopt, or reject, smartphone password managers?
\newblock {\em The 1st European Workshop on Usable Security ({EuroUSEC})},
  2016.

\bibitem{aurigemmaIBISPR2017}
Salvatore Aurigemma, Thomas Mattson, and Lori Leonard.
\newblock So much promise, so little use: What is stopping home end-users from
  using password manager applications?
\newblock {\em The 50th Hawaii International Conference on System Sciences},
  2017.

\bibitem{belenkoBE2012}
Andrey Belenko and Dmitry Sklyarov.
\newblock “secure password managers” and “military-grade encryption” on
  smartphones: Oh, really?
\newblock {\em Blackhat Europe}, page~56, 2012.

\bibitem{caspiAMH2019}
Avner Caspi, Merav Daniel, and Gitit Kav{\'e}.
\newblock Technology makes older adults feel older.
\newblock {\em Aging \& mental health}, 23(8):1025--1030, 2019.

\bibitem{chiassonUSENIX2006}
Sonia Chiasson, Paul~C van Oorschot, and Robert Biddle.
\newblock A usability study and critique of two password managers.
\newblock In {\em USENIX Security Symposium}, volume~15, pages 1--16, 2006.

\bibitem{dasWAY2019}
Sanchari Das, Andrew Kim, Ben Jelen, Joshua Streiff, L~Jean Camp, and Lesa
  Huber.
\newblock Towards implementing inclusive authentication technologies for older
  adults.
\newblock {\em Who Are You}, 2019.

\bibitem{dobranskyPOETICS2016}
Kerry Dobransky and Eszter Hargittai.
\newblock Unrealized potential: Exploring the digital disability divide.
\newblock {\em Poetics}, 58:18--28, 2016.

\bibitem{eluezeABS2018}
Isioma Elueze and Anabel Quan-Haase.
\newblock Privacy attitudes and concerns in the digital lives of older adults:
  Westin’s privacy attitude typology revisited.
\newblock {\em American Behavioral Scientist}, 62(10):1372--1391, 2018.

\bibitem{emami2019exploring}
Pardis Emami-Naeini, Henry Dixon, Yuvraj Agarwal, and Lorrie~Faith Cranor.
\newblock Exploring how privacy and security factor into iot device purchase
  behavior.
\newblock In {\em Proceedings of the 2019 CHI Conference on Human Factors in
  Computing Systems}, pages 1--12, 2019.

\bibitem{faganHCI2017}
Michael Fagan, Yusuf Albayram, Mohammad Maifi~Hasan Khan, and Ross Buck.
\newblock An investigation into users’ considerations towards using password
  managers.
\newblock {\em Human-centric Computing and Information Sciences}, 7(1):12,
  2017.

\bibitem{frikqualitative}
Alisa Frik, Julia Bernd, Noura Alomar, and Serge Egelman.
\newblock A qualitative model of older adults’ contextual decision-making
  about information sharing.
\newblock {\em Workshop on the Economics of Information Security (WEIS 2020)},
  2020.

\bibitem{frikSOUPS2019}
Alisa Frik, Leysan Nurgalieva, Julia Bernd, Joyce Lee, Florian Schaub, and
  Serge Egelman.
\newblock Privacy and security threat models and mitigation strategies of older
  adults.
\newblock In {\em Fifteenth Symposium on Usable Privacy and Security (SOUPS)},
  2019.

\bibitem{gastiESRCS2012}
Paolo Gasti and Kasper~B Rasmussen.
\newblock On the security of password manager database formats.
\newblock In {\em European Symposium on Research in Computer Security}, pages
  770--787. Springer, 2012.

\bibitem{gellGERONTOLOGIST2015}
Nancy~M Gell, Dori~E Rosenberg, George Demiris, Andrea~Z LaCroix, and Kushang~V
  Patel.
\newblock Patterns of technology use among older adults with and without
  disabilities.
\newblock {\em The Gerontologist}, 55(3):412--421, 2015.

\bibitem{haney2018s}
Julie~M Haney and Wayne~G Lutters.
\newblock " it's scary… it's confusing… it's dull": How cybersecurity
  advocates overcome negative perceptions of security.
\newblock In {\em Fourteenth Symposium on Usable Privacy and Security ({SOUPS}
  2018)}, pages 411--425, 2018.

\bibitem{huangAIST2018}
Hsiao-Ying Huang and Masooda Bashir.
\newblock Surfing safely: Examining older adults' online privacy protection
  behaviors.
\newblock {\em Proceedings of the Association for Information Science and
  Technology}, 55(1):188--197, 2018.

\bibitem{ionSOUPS2015}
Iulia Ion, Rob Reeder, and Sunny Consolvo.
\newblock “... no one can hack my mind”: Comparing expert and non-expert
  security practices.
\newblock In {\em Eleventh Symposium On Usable Privacy and Security ({SOUPS}
  2015)}, pages 327--346, 2015.

\bibitem{karoleICISC2010}
Ambarish Karole, Nitesh Saxena, and Nicolas Christin.
\newblock A comparative usability evaluation of traditional password managers.
\newblock In {\em International Conference on Information Security and
  Cryptology}, pages 233--251. Springer, 2010.

\bibitem{linTOCHI2019}
Tian Lin, Daniel~E Capecci, Donovan~M Ellis, Harold~A Rocha, Sandeep Dommaraju,
  Daniela~S Oliveira, and Natalie~C Ebner.
\newblock Susceptibility to spear-phishing emails: Effects of internet user
  demographics and email content.
\newblock {\em ACM Transactions on Computer-Human Interaction (TOCHI)},
  26(5):1--28, 2019.

\bibitem{lyastaniUSENIX2018}
Sanam~Ghorbani Lyastani, Michael Schilling, Sascha Fahl, Michael Backes, and
  Sven Bugiel.
\newblock Better managed than memorized? studying the impact of managers on
  password strength and reuse.
\newblock In {\em 27th {USENIX} Security Symposium ({USENIX} Security 18)},
  pages 203--220, 2018.

\bibitem{nurgalievaPERVASIVE2019}
Leysan Nurgalieva, Alisa Frik, Francesco Ceschel, Serge Egelman, and Maurizio
  Marchese.
\newblock Information design in an aged care context: Views of older adults on
  information sharing in a care triad.
\newblock In {\em Proceedings of the 13th EAI International Conference on
  Pervasive Computing Technologies for Healthcare}, pages 101--110. ACM, 2019.

\bibitem{pearmanSOUPS2019}
Sarah Pearman, Shikun~Aerin Zhang, Lujo Bauer, Nicolas Christin, and
  Lorrie~Faith Cranor.
\newblock Why people (don’t) use password managers effectively.
\newblock In {\em Fifteenth Symposium On Usable Privacy and Security (SOUPS
  2019). USENIX Association, Santa Clara, CA}, pages 319--338, 2019.

\bibitem{poon2020citylab}
Linda Poon and Sarah Holder.
\newblock The ‘new normal’ for many older adults is on the internet.
\newblock {\em Bloomberg CityLab}, May 6, 2020 (last viewed Sept 25, 2020).
\newblock
  \url{https://www.bloomberg.com/news/features/2020-05-06/in-lockdown-seniors\--are-becoming-more-tech-savvy}.

\bibitem{quanhaaseJAIST2020}
Anabel Quan-Haase and Dennis Ho.
\newblock Online privacy concerns and privacy protection strategies among older
  adults in east york, canada.
\newblock {\em Journal of the Association for Information Science and
  Technology}, 2020.

\bibitem{seilerSIGSAC2019}
Sunyoung Seiler-Hwang, Patricia Arias-Cabarcos, Andr{\'e}s Mar{\'\i}n, Florina
  Almenares, Daniel D{\'\i}az-S{\'a}nchez, and Christian Becker.
\newblock " i don't see why i would ever want to use it" analyzing the
  usability of popular smartphone password managers.
\newblock In {\em Proceedings of the 2019 ACM SIGSAC Conference on Computer and
  Communications Security}, pages 1937--1953, 2019.

\bibitem{shayCHI2015}
Richard Shay, Lujo Bauer, Nicolas Christin, Lorrie~Faith Cranor, Alain Forget,
  Saranga Komanduri, Michelle~L Mazurek, William Melicher, Sean~M Segreti, and
  Blase Ur.
\newblock A spoonful of sugar? the impact of guidance and feedback on
  password-creation behavior.
\newblock In {\em Proceedings of the 33rd Annual ACM Conference on Human
  Factors in Computing Systems}, pages 2903--2912, 2015.

\bibitem{silverUSENIX2014}
David Silver, Suman Jana, Dan Boneh, Eric Chen, and Collin Jackson.
\newblock Password managers: Attacks and defenses.
\newblock In {\em 23rd {USENIX} Security Symposium ({USENIX} Security 14)},
  pages 449--464, 2014.

\bibitem{stobertSOUPS2014}
Elizabeth Stobert and Robert Biddle.
\newblock The password life cycle: user behaviour in managing passwords.
\newblock In {\em 10th Symposium On Usable Privacy and Security ({SOUPS}
  2014)}, pages 243--255, 2014.

\bibitem{stobertNSPW2014}
Elizabeth Stobert and Robert Biddle.
\newblock A password manager that doesn't remember passwords.
\newblock In {\em Proceedings of the 2014 New Security Paradigms Workshop},
  pages 39--52, 2014.

\bibitem{tsaiEG2015}
Hsin-yi~Sandy Tsai, Ruth Shillair, Shelia~R Cotten, Vicki Winstead, and
  Elizabeth Yost.
\newblock Getting grandma online: are tablets the answer for increasing digital
  inclusion for older adults in the us?
\newblock {\em Educational gerontology}, 41(10):695--709, 2015.

\bibitem{urUSENIX2012}
Blase Ur, Patrick~Gage Kelley, Saranga Komanduri, Joel Lee, Michael Maass,
  Michelle~L Mazurek, Timothy Passaro, Richard Shay, Timothy Vidas, Lujo Bauer,
  et~al.
\newblock How does your password measure up? the effect of strength meters on
  password creation.
\newblock In {\em Presented as part of the 21st {USENIX} Security Symposium
  ({USENIX} Security 12)}, pages 65--80, 2012.

\bibitem{urUSENIX2015}
Blase Ur, Fumiko Noma, Jonathan Bees, Sean~M Segreti, Richard Shay, Lujo Bauer,
  Nicolas Christin, and Lorrie~Faith Cranor.
\newblock " i added'!'at the end to make it secure": Observing password
  creation in the lab.
\newblock In {\em Eleventh Symposium On Usable Privacy and Security ({SOUPS}
  2015)}, pages 123--140, 2015.

\bibitem{washSOUPS2016}
Rick Wash, Emilee Rader, Ruthie Berman, and Zac Wellmer.
\newblock Understanding password choices: How frequently entered passwords are
  re-used across websites.
\newblock In {\em Twelfth Symposium on Usable Privacy and Security ({SOUPS}
  2016)}, pages 175--188, 2016.

\bibitem{wolf2019pretty}
Flynn Wolf, Ravi Kuber, and Adam~J Aviv.
\newblock " pretty close to a must-have" balancing usability desire and
  security concern in biometric adoption.
\newblock In {\em Proceedings of the 2019 CHI Conference on Human Factors in
  Computing Systems}, pages 1--12, 2019.

\end{thebibliography}


\appendix
\section*{Appendix}

We used an interview script identical to that of Pearman et al.~\cite{pearmanSOUPS2019}. The codebook used by Pearman et al. is available at \url{https://osf.io/6u7m8/}. 
\bigskip



\noindent{\bf General Questions about Passwords}
\begin{enumerate}[noitemsep]
    \item What types of online accounts do you have? (e.g. social media, bank accounts, shopping sites, etc.)
    \item What level of protection do you think they each need?(Follow up, if necessary): Are there some accounts you want to protect more than others?
    \item To the best of your knowledge, approximately how many online accounts do you have that use passwords?
    \item How many of these do you access on a daily basis?
    \item On which device(s) do you access these online accounts? Follow-up below for each category the person has.)
    \begin{enumerate}[noitemsep]
        \item For phones/tablets: what type(s)? (iPhone, Android,etc.)
        \item For computers: what operating system(s)? (Windows, Mac, Linux, ChromeOS, etc.)
        \item Public, work or personal device?
        \item For each device:  what web browser do you use most often on your [device]?
    \end{enumerate}
    \item How many times do you manually type in passwords on a daily basis? (Which types of accounts? On which device(s)?)
    \item How many of your accounts are always logged in? (Which types of accounts? On which device(s)?)
    \item Do you have any passwords that get auto-filled for you?
    (Which types of accounts? On which device(s)? Do you know how your passwords are auto-filled)
    \item  Are your passwords different for each account?
    \begin{enumerate}[noitemsep]
        \item (If yes) Are your passwords similar to one another?
        \item (if reuse exists): How many of your accounts share the same password? How many of your accounts have unique passwords?
    \end{enumerate}
    \item How do you create a password for a new account?
    \begin{enumerate}[noitemsep]
        \item How does this password compare to other pass-words? (i.e. is it similar?)
        \item What if your password does not meet the character/length requirements. How would you change your password to meet those requirements?
        \item Is this process different for some types of accounts?Which ones? What do you do?        
    \end{enumerate}
    \item How do you keep track of your passwords now? Do you use more than one method?
    \item Are you satisfied with your current method(s) of managing your passwords? (What do you find easy about it? What do you find difficult about it?)
    \item Has anyone ever logged into any of your accounts with-out your permission?
    \begin{enumerate}[noitemsep]
        \item (if  yes)  Was  this  done  by  someone  you  didn't know? 
        \item (if yes) What did you do? Follow up, if applicable: Did you change the compromised password? How did you choose the new password? How does the new password compare to your existing passwords? Did you change the passwords to your other accounts that share the same password?
        \item (if no) What would you do if someone did? (Would  you  change  the  compromised  pass-word? If yes, how would you choose the new password? How would you choose it?
    \end{enumerate}
    \item   To your knowledge, have any of your accounts ever been subject to a password data breach?
    \begin{enumerate}
        \item (if yes) How did you find out about it? What did you do? After the breach, did you change the way you manage your passwords? Did that account share a password with any of your other accounts? If so, did you change any of those passwords?
        \item (if no) What would you do if it was?
    \end{enumerate}
\end{enumerate}
\smallskip{}

\noindent{\bf General Questions about Passwords Managers}
\begin{enumerate}[noitemsep]
    \item Have you ever heard of password managers? Where didyou hear about them?
    \item Do you use a password manager?
    \item What, to your knowledge, is the purpose of a password manager? (If they respond to something along the lines of "it manages  passwords")  What else  do  you  think they're used for?
    \item {\em Read description of password managers to participant}
    \smallskip
    
    Password managers are tools that can securely handle passwords for you. They can remember your passwords,generate new ones, and even sync them across devices.There are various types of password managers with different features, but for the purpose of this interview, we will consider three of them.
    
    One  type  of password manager is  built into  the  web browser, such as Google Chrome, Mozilla Firefox, Safari,Internet Explorer, and Microsoft Edge. These browser scan remember passwords for websites, as well as autofill them for you
    
    Another type of password manager is a third-party application. This can be software you install directly onto your devices or a service you can access on the web. It can also remember and/or autofill your passwords, including across browsers and devices.
    
    Lastly, your operating system can serve as a password manager as well. For example, the Keychain functionality on MacOS can remember passwords in and out of your browser. It can also be used with iCloud to sync passwords across Apple devices.
    
    Ultimately, the main purpose of password managers is to automatically handle your passwords for you.
    
    \item Based on our description, which of these categories ofpassword managers do you currently use, if any?
    
    \item Have you used any [other] password manager tools in the past?
    \item (If they have used PM, now or in the past) When did you start using a password manager? Why did you start using it?
    \item (if stopped use): When did you stop using the password manager and why?
    \item (If they use any and haven’t already named them) Can you name the password management tools that you use?(Or if they can’t name them, ask them to describe them /indicate how they use them so that you can try to discern what they mean)
\end{enumerate}
\smallskip

\noindent{\bf Experience using Password Managers}
\begin{enumerate}[noitemsep]
    \item Why did you choose [PM]?
    \item How has your experience been using a password manager?
    \item What functions did you like / find helpful?
    \item What functions did you dislike / find unhelpful?
    \item Is all functionality of your password manager available for free, or does this tool have a paid version?
    \begin{enumerate}[noitemsep]
        \item (If paid version exists) Do you use the paid or free version? Why?
        \item  (if uses free version) Would you ever pay for a password manager?  How much?  What features would it have?
    \end{enumerate}
    \item Do you use your password manager on all of your devices, including [list of tools they already told you about in the first section] (if no, which devices do you use it on? Why do you use it on those? Why not use it on the others? How do you keep track of passwords on the device(s) that you don’t use your PM on?)
    \item (For each device that the user uses PM on): Did you have to install an application to your device, or install an extension to your browser, or both? (if no, How do you access your password manager? Possible answers include logging into a website,or USB drive)
    \item Does your password manager offer the option of syncing passwords between devices? (If this option exists, do you use it? Why or why not?)
    \item Do you use your password manager for all the accounts you access through your web browser? (If not, how do you decide which accounts to use it for? How do you keep track of passwords that are not stored in this PM?)
    \item Do you use your password manager for any accounts outside of your web browser? Examples of this would include an email client like Outlook on your computer or a social media app such as Facebook on your phone.
    \begin{enumerate}[noitemsep]
        \item Do you use it for all of the accounts outside of your web browser(s)?
        \item (if no to a) How do you decide which accounts to use it for?
    \end{enumerate}
    \item Do you have to provide a master password or other authentication to access the passwords stored in your pass-word manager?
    \begin{enumerate}
        \item (If yes) What type of password or authentication is required?
        \begin{enumerate}[noitemsep]
            \item (if master password): How  did  you  create  your  master  password? Is your master password similar to your other passwords? Is  it  difficult to  remember your master password? (if yes) How do you remember it?        
        \end{enumerate}
        \item (If yes) How often do you have to provide it?
        \item Have  you  ever modified  the  default  settings  to change how often you have to provide this?
    \end{enumerate}
    \item Do you feel like your passwords are safe and secure when stored in this PM tool?
    \item Do you know how this tool protects the security of your passwords? {\em (Unless they say they have no idea, ask them to elaborate on how they think it works)}
    \item Does your password manager have a password generation
    tool?
    \begin{enumerate}[noitemsep]
        \item (if yes) Have you ever used the password generation tool? (if yes, below)
        \begin{enumerate}[noitemsep]
        \item Do you use the generation tool for newly created accounts?
        \item Have you used the tool to generate a new password for an existing account?
        \item (if yes to B) Does your password manage have an automatic password replacement feature that changes passwords for you without you having to actually visit the website yourself? Do you use it? Why or why not?
        \item Approximately how many of your passwords are now created by the password generation functionality?
        \item Do you ever change the settings from the defaults when generating a password?
        \item Was there an instance where the generated password did not meet the website's password requirements? (If yes) What did you do about it?
        \item Overall, how has your experience been using the password generation tool?
        \end{enumerate}
    \end{enumerate}
    \item Does your password manager have a dashboard or tool hat examines the security of your passwords?
    \begin{enumerate}
        \item (If yes): How often do you use it?
        \item Have  you  changed any of your passwords  after looking at this information?
    \end{enumerate}
    \item Has your password manager ever informed you of a data breach? (If yes, what did you do?
    \item Has  your  password  manager  ever  prompted  you  to change your password? (if yes, under what situation? What did you do?)
    \item Are there any additional services or features that you would want in your password manager?
\end{enumerate}

\noindent{\bf Why not Using PMs? (If answer “no” to Using Password Managers)}
\begin{enumerate}[noitemsep]
    \item Can you tell us why you aren't using a password manager? {\em (Follow up by probing what it would take for them to use a password manager.)}
    \item Many third-party password managers require a monthly fee to use their services. Would you be willing to pay for such a service?
    \begin{enumerate}[noitemsep]
        \item If yes, how much?
        \item If no, why not? ({\em If participant says there are free third-party PMs available, then ask:} Would you be willing to pay for additional features that are not included in the free version? How much would you be willing to pay?)
    \end{enumerate}
\end{enumerate}

\noindent{\bf Perceptions of Password Managers’ Functions}

We talked about different types of password managers a few minutes ago, including third-party password managers, pass-word managers built into web browsers, and password managers built into operating systems.
\begin{enumerate}[noitemsep]
    \item Do you think some types of password manager tools are safer to use than others? (Why?)
    \item Do you think some types of password manager tools are more convenient than others? (Why?)
    \item How do you think password manager tools compare to other methods of managing passwords, such as writing them down on paper or saving them in a file on your computer? (Why?)
    \item How do you think password managers store passwords?
    \begin{enumerate}[noitemsep]
        \item Do you think password managers store your pass-words locally on your device or on a server (in the cloud)? Do  you  think one  is  more  secure  than  the other? (If so, which one? Why?) Do you have a preference? Why or why not?
        \item How do you think password managers sync your accounts across devices? Would you want this function? Why? Do you think this impacts your password security? If so, how?
        \item What do you think the password data looks like when stored on your computer? If your password is "password2018!", does your  password  manager  store  it  as  "password2018!"? Is there a difference when stored in the cloud?
    \end{enumerate}
    \item Do you think password managers affect the security of your accounts? Why or why not?
    \item Do you trust password managers to always store or not forget your passwords? Why or why not?
    \item Do you trust password managers to protect your passwords from attackers? Why or why not?
    \item Have you ever received advice or training on how to create or manage passwords? 
    \begin{enumerate}[noitemsep]
        \item (if yes) What guidelines have you been taught? Where?
        \item (if yes) Do you use these guidelines? Why or why not?
    \end{enumerate}
    \item (non-PM user): Would you consider using a password manager in the future? Why or why not?
    \item (If "No" or "I don’t know" to data breach question) earlier you mentioned that you were never impacted by a data breach, or that you weren't sure if you were. Would you like the opportunity to verify this?We can use a website called HaveIBeenPwned to check whether your accounts were compromised in a public data breach.
    \begin{enumerate}[noitemsep]
        \item  Explain to participant:The website asks for your email address and checks if any accounts tied to it were compromised. Note, however, that the website cannot check information on every data breach. It checks those that are known to the public
        \item If participant agrees, inform participant:For privacy reasons, we recommend that you access the website on your own device. This way, we won't see your email address, nor will we know which of your accounts, if any, were impacted by a breach.
        \item Instruct the participant to try any other email address they may use often.
        \item Were any of your accounts compromised? (if yes, below)
        \begin{enumerate}[noitemsep]
            \item How many?
            \item What types of accounts? (Provide categories to choose from: social media, bank, shopping,other)
            \item How do you feel about this information?
            \item (follow up, if necessary) Will you do anythingwith this information?
        \end{enumerate}
    \end{enumerate}
\end{enumerate}

\end{document}